\documentclass[american,aps,pra,reprint,superscriptaddress]{revtex4-1}
\usepackage[T1]{fontenc}
\usepackage[latin9]{inputenc}
\usepackage{babel}
\usepackage{amsmath}
\usepackage{amsthm}
\usepackage{amssymb}
\usepackage{graphicx}
\usepackage[unicode=true,pdfusetitle,
 bookmarks=true,bookmarksnumbered=false,bookmarksopen=false,
 breaklinks=false,pdfborder={0 0 0},backref=false,colorlinks=false]
 {hyperref}
\hypersetup{
 colorlinks,linkcolor=myurlcolor,citecolor=myurlcolor,urlcolor=myurlcolor}

\makeatletter
\theoremstyle{plain}
\newtheorem{thm}{\protect\theoremname}
  \theoremstyle{plain}
  \newtheorem{prop}[thm]{\protect\propositionname}

\usepackage{babel}
\usepackage{babel}
\usepackage{babel}
\usepackage{babel}
\usepackage{babel}
\usepackage{txfonts}
\usepackage{colortbl}\definecolor{myurlcolor}{rgb}{0,0,0.7}
\newcommand{\tr}{{\operatorname{Tr\,}}}
\newcommand{\id}{{\operatorname{id}}}

\def\ket#1{| #1 \rangle}
\def\bra#1{\langle  #1 |}
\def\braket#1{\langle  #1 \rangle}
\def\proj#1{| #1 \rangle\!\langle #1 |}

\providecommand{\propositionname}{Proposition}
\providecommand{\theoremname}{Theorem}

\providecommand{\propositionname}{Proposition}
\providecommand{\theoremname}{Theorem}

\providecommand{\propositionname}{Proposition}
\providecommand{\theoremname}{Theorem}

\providecommand{\propositionname}{Proposition}
\providecommand{\theoremname}{Theorem}

  \providecommand{\propositionname}{Proposition}
\providecommand{\theoremname}{Theorem}

\makeatother

  \providecommand{\propositionname}{Proposition}
\providecommand{\theoremname}{Theorem}

\begin{document}

\title{Entanglement and coherence in quantum state merging}

\author{A. Streltsov}

\email{streltsov.physics@gmail.com}

\affiliation{ICFO -- Institut de Ciències Fotòniques, The Barcelona Institute
of Science and Technology, ES-08860 Castelldefels, Spain}

\affiliation{Dahlem Center for Complex Quantum Systems, Freie Universität Berlin,
D-14195 Berlin, Germany}

\author{E. Chitambar}

\affiliation{Department of Physics and Astronomy, Southern Illinois University,
Carbondale, Illinois 62901, USA}

\author{S. Rana}

\affiliation{ICFO -- Institut de Ciències Fotòniques, The Barcelona Institute
of Science and Technology, ES-08860 Castelldefels, Spain}

\author{M. N. Bera}

\affiliation{ICFO -- Institut de Ciències Fotòniques, The Barcelona Institute
of Science and Technology, ES-08860 Castelldefels, Spain}

\author{A. Winter}

\affiliation{F\'{i}sica Teòrica: Informació i Fenòmens Quàntics, Universitat Autònoma
de Barcelona, ES-08193 Bellaterra (Barcelona), Spain}

\affiliation{ICREA -- Institució Catalana de Recerca i Estudis Avançats, Pg.~Lluis
Companys 23, ES-08010 Barcelona, Spain}

\author{M. Lewenstein}

\affiliation{ICFO -- Institut de Ciències Fotòniques, The Barcelona Institute
of Science and Technology, ES-08860 Castelldefels, Spain}

\affiliation{ICREA -- Institució Catalana de Recerca i Estudis Avançats, Pg.~Lluis
Companys 23, ES-08010 Barcelona, Spain}
\begin{abstract}
Understanding the resource consumption in distributed scenarios is
one of the main goals of quantum information theory. A prominent example
for such a scenario is the task of quantum state merging where two
parties aim to merge their parts of a tripartite quantum state. In
standard quantum state merging, entanglement is considered as an expensive
resource, while local quantum operations can be performed at no additional
cost. However, recent developments show that some local operations
could be more expensive than others: it is reasonable to distinguish
between local incoherent operations and local operations which can
create coherence. This idea leads us to the task of incoherent quantum
state merging, where one of the parties has free access to local incoherent
operations only. In this case the resources of the process are quantified
by pairs of entanglement and coherence. Here, we develop tools for
studying this process, and apply them to several relevant scenarios.
While quantum state merging can lead to a gain of entanglement, our
results imply that no merging procedure can gain entanglement and
coherence at the same time. We also provide a general lower bound
on the entanglement-coherence sum, and show that the bound is tight
for all pure states. Our results also lead to an incoherent version
of Schumacher compression: in this case the compression rate is equal
to the von Neumann entropy of the diagonal elements of the corresponding
quantum state. 
\end{abstract}
\maketitle
\textbf{\emph{Introduction}}\textbf{.} While coherence has long been
known in classical physics as a fundamental waves property \cite{Young1807},
in quantum mechanics coherent superposition is elevated to a universal
principle governing all processes. Indeed, the fact that all matter
exhibits wave behavior was first understood by de Broglie \cite{DeBroglie1922,*DeBroglie1923,*DeBroglie1924},
which became the basis of the now standard formulation of quantum
mechanics in Schrödinger's wave equation \cite{Schrodinger1926a,*Schrodinger1926b,*Schrodinger1926e,*Schrodinger1926c,*Schrodinger1926d,*Schrodinger1926f,*Schrodinger1926g}.
The universality of the superposition principle, i.e. the \emph{tenet}
that any two valid states of a system can be superposed to form a
new valid state, marks a radical departure from classical physics.
It is at the heart of the many counterintuitive features of quantum
theory, perhaps most famously in Schrödinger's Gedankenexperiment
of the cat \cite{Schrodinger1935a,*Schrodinger1935b,*Schrodinger1935c}.
Quantum entanglement can be considered as a particular manifestation
of coherence, and both of these nonclassical phenomena have led to
extensive debates in the early days of quantum mechanics \cite{Einstein1935,Bohr1935}.

While the study of the resource theory of entanglement has a long
tradition \cite{Plenio2007,Horodecki2009}, the resource theory of
quantum coherence has been formulated only recently \cite{Baumgratz2014,Winter2015},
although other attempts in this direction have been also presented
earlier \cite{Aberg2006,Gour2008,Gour2009,Levi14,Marvian2013,Marvian2014}.
The basis of any resource theory are free states, these are states
which can be created at no cost. In entanglement theory, these are
all separable states. In coherence theory these are incoherent states
\cite{Baumgratz2014}, i.e., states which are diagonal in a fixed
basis $\ket{i}$. The second important ingredient of any resource
theory are free operations, i.e., operations which can be performed
at no additional cost. In entanglement theory this is usually the
set of local operations and classical communication, although other
more general sets such as separable operations \cite{Rains1997,Vedral1998}
and asymptotically nonentangling operations \cite{Brandao2008,Brandao2010}
have also been considered. In coherence theory, free operations are
called incoherent operations. These are precisely the quantum operations
which have incoherent Kraus operators, i.e., $K_{i}\ket{m}\propto\ket{n}$,
where $\ket{m}$ and $\ket{n}$ are elements of the incoherent basis
\cite{Baumgratz2014}.

Triggered by these recent developments, much effort is put into understanding
the role of coherence as a resource in quantum theory \cite{Chen2015,Cheng2015,Mondal2015,Deb2015,Du2015b,Bera2015,Liu2015,Zou2014,Prillwitz2014,Du2015a,Hillery2015,Yadin2015b,Bu2015b,Matera2015,Singh2015b,Zhang2015,Singh2015a,Chen2016}.
Several new quantifiers of coherence have been proposed \cite{Girolami2014,Du2015,Yuan2015,Pires2015,Qi2015,Xu2015,Yadin2015,Rana2016,Allegra2015,Chandrashekar2016,Chitambar2016,Streltsov2015e,Napoli2016,Piani2016},
and the dynamics of some of these quantities under noisy evolution
has been investigated \cite{Bromley2015,Singh2015,Chanda2015,Zhang2015a,Silva2015}.
Several works also study maximally coherent states \cite{Peng2015,Bai2015},
the role of coherence in spin models \cite{Karpat2014,Cakmak2015},
cohering power of quantum channels \cite{Mani2015,Bu2015,Garcia-Diaz2015},
and relations between coherence and other measures of quantumness
\cite{Streltsov2015b,Killoran2015,Ma2016,Yao2015,Xi2014,Hu2015,Girolami2015}.
Coherence also plays an important role in quantum thermodynamics \cite{Skrzypczyk2014,Aberg2014,Lostaglio2015a,Lostaglio2015b,Korzekwa2015,Kammerlander2015,Vacanti2015,Gour2013,Malvezzi2016,Bhattacharya2016,Yang2015},
and its investigation in biological systems is an important step towards
finding quantum phenomena in living objects \cite{Lloyd2011,Li2012,Huelga13,Singh2014}.
Additionally, a distinction between ``speakable'' and ``unspeakable''
coherence has also been introduced recently \cite{Marvian2016}. Here
we are describing coherence in a speakable sense whereas unspeakable
coherence is the resource captured in resource theories of asymmetry
\cite{Marvian2013}.

Contrary to entanglement, which inherently implies a scenario of at
least two separated parties, the resource theory of coherence has
been initially introduced for one party only. Very recently, there
were several approaches to extend the notion of coherence to more
than one party \cite{Bromley2015,Streltsov2015b,Yao2015,Hu2015a,Chitambar2015,Chitambar2015a,Streltsov2015c,Kumar2015,Hu2015,Mondal2015a}.
Here, we build on the methods presented in \cite{Chitambar2015,Streltsov2015c,Chitambar2015a},
aiming to study the interplay between entanglement and coherence in
the task known as quantum state merging \cite{Horodecki2005a,Horodecki2007}.

In standard quantum state merging, two parties -- their names are
traditionally Alice and Bob -- share a mixed quantum state. Alice
aims to send her part of the state to Bob via an additional quantum
channel. The difficulty of the task arises from an extra requirement:
the process has to be performed in such a way that the overall purification
of the state remains intact. As was shown in \cite{Horodecki2005a,Horodecki2007},
the singlet rate required for this process is equal to the conditional
entropy $S(\rho^{AB})-S(\rho^{B})$, where $S(\rho)=-\tr[\rho\log_{2}\rho]$
is the von Neumann entropy. To be precise, if the conditional entropy
is positive, then merging is possible with singlets at rate $S(\rho^{AB})-S(\rho^{B})$,
and merging is not possible if less singlets are available. Moreover,
if the conditional entropy is negative, the process is possible without
any entanglement. Apart from merging the state for free, Alice and
Bob can additionally gain singlets at rate $S(\rho^{B})-S(\rho^{AB})$.

Here, we consider the task of \emph{incoherent quantum state merging}.
This task is very similar to standard quantum state merging, up to
the fact that Bob has free access to incoherent operations only, i.e.,
he has to pay for operations which are not incoherent. There are at
least two motivations for this: On the one hand, we would like to
understand better the local quantum(!) operations that Alice and in
particular Bob have to perform in merging. On the other hand, coherence
seems to be the resource of choice to consider here, as entanglement
and coherence are both resources of superposition, one in correlation,
the other locally. Thus, while the cost of standard quantum state
merging is quantified by the required entanglement rate $E$, the
cost of incoherent quantum state merging will be quantified by a pair
of entanglement and coherence rate $(E,C)$. Solving the problem of
incoherent quantum state merging requires the characterization of
all \emph{optimal pairs} $(E,C)$. These are pairs of entanglement
and coherence for which merging is possible, but neither entanglement
nor coherence of the pair can be reduced.

In this paper we define the task of incoherent quantum state merging
and develop methods to study it. For arbitrary mixed states we provide
a powerful lower bound on the entanglement-coherence sum $E+C$. For
pure states we show that this bound is tight by explicitly evaluating
the minimal singlet rate needed for merging in the absence of local
coherence. For a family of fully separable mixed states we solve the
question of incoherent quantum state merging completely by presenting
all optimal pairs of entanglement and coherence. Finally, we provide
a discussion on the interplay between coherence and entanglement,
also presenting evidence that a large amount of local coherence might
be saved by using little extra entanglement in the merging procedure.

At this point we note that the term \textquotedbl{}coherence\textquotedbl{}
used in this and other recent papers is, of course, also used in atomic
and molecular physics, where \textquotedbl{}coherences\textquotedbl{}
denote off-diagonal elements of the density matrix, typically in the
basis of energy eigenstates. Note, however, that in quantum optics
the term \textquotedbl{}coherence\textquotedbl{} is also used in the
context on classical and quantum electrodynamics, where it describes
the factorization property of certain correlation functions, ultimately
related to the prominent Glauber-Sudarshan \textquotedbl{}coherent
states\textquotedbl{} \cite{Glauber1963,Sudarshan63}. Off-diagonal
elements of the density matrix in this latter sense, are related rather
to \textquotedbl{}non-classicality\textquotedbl{} of states of photos,
phonons, bosons etc. (cf. \cite{Korbicz2005,Vogel2006,Schleich2001}
and references therein).

\bigskip{}

\textbf{\emph{Incoherent quantum state merging.}} We consider the
scenario where three parties, Alice, Bob, and a referee, share a joint
quantum state $\rho=\rho^{RAB}$. In the task of incoherent quantum
state merging, Alice and Bob aim to merge their parts of the total
state on Bob's side by using local quantum-incoherent operations and
classical communication (LQICC) \cite{Chitambar2015}. Additionally,
Alice and Bob have access to singlets at rate $E$ and maximally coherent
states at rate $C$ on Bob's side.

In the following, we are interested in \emph{achievable pairs} $(E,C)$,
these are pairs of coherence and entanglement for which the aforementioned
task can be performed in the asymptotic scenario. Similar to standard
quantum state merging \cite{Horodecki2005a,Horodecki2007} we consider
the most general situation, where Alice and Bob can make catalytic
use of entanglement and coherence \footnote{In general, a process makes catalytic use of a resource if it starts
with some amount of the resource $R_{1}$ and ends with some nonzero
amount $R_{2}>0$.}. We call $E_{i}$ the entanglement rate which is initially shared
by Alice and Bob, and $E_{t}$ will be the final amount of entanglement
between them. Similarly, $C_{i}$ and $C_{t}$ will be the initial
and the final amount of Bob's local coherence. An entanglement-coherence
pair $(E,C)$ is achievable if there exist numbers $E_{i}$, $E_{t}$,
$C_{i}$, and $C_{t}$ with $E=E_{i}-E_{t}$ and $C=C_{i}-C_{t}$
such that for any $\varepsilon>0$ and any $\delta>0$ for all sufficiently
large integers $n\geq n_{0}$ there exists an LQICC protocol $\Lambda$
between Alice and Bob such that \footnote{It is easy to see that there exists at least one achievable pair.
In particular, if Alice's system has dimension $d_{A}=2^{n}$, then
an achievable pair is given by $(E=n,C=0)$. This follows from the
fact that single-qubit teleportation can be performed with one singlet
and without additional coherence \cite{Streltsov2015c}. If the dimension
of Alice's system does not have this form, there still exists an integer
$m$ such that $2^{m}>d_{A}$. Then, $(E=m,C=0)$ is an achievable
pair by the same arguments.} 
\begin{equation}
\left\Vert \Lambda\!\left[\rho_{i}^{\otimes n}\!\otimes\!\Phi_{2}^{\otimes\lfloor(E_{i}+\delta)n\rfloor}\!\otimes\!\Psi_{2}^{\otimes\lfloor(C_{i}\text{+\ensuremath{\delta})}n\rfloor}\right]-\rho_{t}^{\otimes n}\!\otimes\!\Phi_{2}^{\otimes\lceil E_{t}n\rceil}\!\otimes\!\Psi_{2}^{\otimes\lceil C_{t}n\rceil}\right\Vert _{1}\!\leq\varepsilon.\label{eq:achievable}
\end{equation}
Here, $\rho_{i}=\rho^{RAB}\otimes\proj{0}^{\tilde{B}}$ is the total
initial state, where $\tilde{B}$ is an additional particle in Bob's
hands with dimension $d_{\tilde{B}}=d_{A}$. $\ket{\Phi_{2}}=\sqrt{\frac{1}{2}}(\ket{00}+\ket{11})$
is a maximally entangled two-qubit state shared by Alice and Bob,
and $\ket{\Psi_{2}}=\sqrt{\frac{1}{2}}(\ket{0}+\ket{1})$ is a maximally
coherent single-qubit state on Bob's side. The target state $\rho_{t}=\rho^{R\tilde{B}B}\otimes\proj{0}^{A}$
is the same as $\rho_{i}$ up to relabeling the parties $A$ and $\tilde{B}$,
and $\|M\|_{1}=\tr\sqrt{M^{\dagger}M}$ is the trace norm.

The achievable region is a closed and convex set, due to the timesharing
principle \cite{Cover2006,Csiszar2011}. Namely, on block length $n$,
and for $0<p<1$, we can break the $n$ systems into two blocks of
$k=\lfloor pn\rfloor$ and $\ell=\lceil(1-p)n\rceil$, and run a first
protocol with asymptotic rate $(E_{1},C_{1})$ on the $k$-block,
and a second protocol with asymptotic rate $(E_{2},C_{2})$ on the
$\ell$-block. The tensor product of these protocols is evidently
an asymptotically error-free merging protocol, and achieves the rate
pair $(E,C)=(pE_{1}+(1-p)E_{2},pC_{1}+(1-p)C_{2})$.

As in standard quantum state merging, the quantities $E$ and $C$
can be positive or negative. If $E$ ($C$) is positive, it means
that the merging procedure consumes entanglement (coherence) at rate
$E$ ($C$). If the corresponding quantity is negative, the process
can be performed without the corresponding resource, and additionally
singlets (maximally coherent states) are gained. Crucially, as we
will see below in this paper, the latter gain is not possible for
both entanglement and coherence at the same time: if entanglement
is gained in the process, coherence has to be consumed, and vice versa.

Clearly, if a pair $(E,C)$ is achievable, then any other pair $(E',C')$
is also achievable for $E'\geq E$ and $C'\geq C$. A pair $(E,C)$
will be called \emph{optimal} if it is achievable and if the pairs
$(E,C')$ and $(E',C)$ are not achievable for any $C'<C$ and $E'<E$.
Since via LQICC operations a singlet can be converted into a maximally
coherent state on Bob's side \cite{Chitambar2015}, with every achievable
pair $(E,C)$, also $(E+t,C-t)$ is achievable for $t>0$. Thus, it
is always possible to perform incoherent merging with $C=0$, and
the corresponding optimal pair will be denoted $(E_{0},0)$. Another
important pair is the one with the minimal amount of entanglement
$E_{\min}$ among all protocols. We denote it $(E_{\min},C_{\max})$,
since it also has the maximal amount of coherence among all optimal
pairs \footnote{To see that the pair $(E_{\min},C_{\max})$ has maximal amount of
coherence among all optimal pairs, consider two optimal pairs $(E,C)$
and $(E',C')$ with $C<C'$. Then, by optimality it must be that $E>E'$.
Suppose now, there existed a pair $(\tilde{E},\tilde{C})$ with $\tilde{C}>C_{\max}$.
By the aforementioned argument it follows that $\tilde{E}<E_{\min}$,
which is a contradiction.}.

A full solution of incoherent quantum state merging implies determining
all optimal pairs for a given tripartite state. The following proposition
provides a bound on the entanglement-coherence sum $E+C$. 
\begin{prop}
\label{prop:bound-1}Given a tripartite quantum state $\rho=\rho^{RAB}$,
any achievable pair $(E,C)$ fulfills the following inequality: 
\begin{equation}
E+C\geq S\left(\id^{R}\otimes\Delta^{AB}\left[\rho\right]\right)-S\left(\id^{RA}\otimes\Delta^{B}\left[\rho\right]\right),\label{eq:bound-1}
\end{equation}
where $\Delta^{X}[\rho]$ denotes full decoherence of the state $\rho$
in the incoherent basis of a (possibly multipartite) subsystem $X$:
\begin{equation}
\Delta^{X}[\rho]=\sum_{i}\proj{i}^{X}\rho\proj{i}^{X}.
\end{equation}

\end{prop}
\noindent We refer the reader to Appendix \ref{sub:Proposition1}
for the proof, which is based on monotonicity of QI relative entropy
under LQICC operations \cite{Chitambar2015}.

It is instructive to compare these results to standard quantum state
merging as presented in \cite{Horodecki2005a,Horodecki2007}. In standard
quantum state merging, the entanglement rate required for merging
a pure state $\psi^{RAB}$ is given by the conditional entropy of
the reduced state $\rho^{AB}$, which can be either positive or negative.
In the negative case, quantum state merging is possible without entanglement
and additional singlets are produced. Since the right-hand side of
Eq.~(\ref{eq:bound-1}) cannot be negative, it follows that the sum
$E+C$ is also nonnegative. While each of the quantities $E$ or $C$
can still be negative individually, they cannot be both negative at
the same time. Thus, there is no merging procedure where entanglement
and coherence are gained simultaneously. This statement is true for
all mixed states $\rho^{RAB}$.

Having presented the general framework, we will now focus on the situation
where the total state is pure. Note that understanding of the pure-state
scenario also gives insights for general mixed states. In particular,
if a pair $(E,C)$ is achievable for a pure state $\ket{\psi}^{RAB}$,
the same pair is also achievable for any state $\rho^{RAB}$ with
the same reduction such that $\rho^{AB}=\mathrm{Tr}_{R}[\psi^{RAB}]$.

\bigskip{}

\textbf{\emph{Incoherent merging of pure states.}} We will now consider
incoherent quantum state merging for general pure states. By state
merging \cite{Horodecki2005,Horodecki2007} we have 
\begin{equation}
E\geq E_{\min}=S(\rho^{AB})-S(\rho^{B}).
\end{equation}
Moreover, for pure states Proposition~\ref{prop:bound-1} reduces
to 
\begin{equation}
E+C\geq S(\overline{\rho}^{AB})-S(\overline{\rho}^{B}),
\end{equation}
where we introduced the notation $\overline{\rho}^{X}=\Delta^{X}[\rho^{X}]$
for the dephased state. As we will see in the following theorem, this
bound is saturated. 
\begin{thm}
\label{thm:pure-merging}Any pure state $\ket{\psi}^{RAB}$ can be
merged with the pair $(E_{0},C=0)$, where 
\begin{equation}
E_{0}=S\left(\overline{\rho}^{AB}\right)-S\left(\overline{\rho}^{B}\right).
\end{equation}
Moreover, the pair $(E_{0},C=0)$ is optimal. 
\end{thm}
\noindent We refer to Appendix \ref{sec:Theorem2} for the proof,
which is based on an adaptation of the Slepian-Wolf distributed compression
of the decohered - classical! - source. Note that $\overline{\rho}^{AB}$
is a classical state, and its conditional entropy, according to the
Slepian-Wolf theorem \cite{Slepian1973}, is precisely the amount
of classical communication required to inform Bob about Alice's register.
In fact, the proof of this theorem in Appendix \ref{sec:Theorem2}
uses the Slepian-Wolf protocol as a building block.

The above theorem implies that for pure states $\psi^{RAB}$ the minimal
entanglement-coherence sum $E+C$ required for merging is equal to
the conditional entropy of the decohered state $\overline{\rho}^{AB}$.
We also mention that for pure states of the form $\ket{\psi}^{RA}\otimes\ket{0}^{B}$,
the procedure described here can be seen as the incoherent version
of Schumacher compression \cite{Schumacher1995}. In particular, Theorem
\ref{thm:pure-merging} proves that any state $\rho$ can be faithfully
compressed at rate $S(\Delta[\rho])$, under the assumption that the
decompression is performed with incoherent operations only.

A final comment is in order concerning the applicability of Proposition
\ref{prop:bound-1} and Theorem \ref{thm:pure-merging} to different
operational classes. Beyond the incoherent operations considered in
this letter, one can consider the more general class of ``maximal''
incoherent operations (MIO), which consists of all non-coherence-generating
maps \cite{Brandao2015,Winter2015}. As we discuss in Appendix \ref{sub:Proposition1},
the lower bound of \ref{prop:bound-1} holds as well for MIO. On the
achievability end, the rate of Theorem \ref{thm:pure-merging} is
still achievable when Bob is limited to so-called \textit{strictly}
incoherent operations (SIO) \cite{Winter2015,Yadin2015b}, and even
if he is further restricted to the class of \textit{physical} incoherent
operations (PIO) \cite{Chitambar2016}. Also, Alice's measurement
in Theorem \ref{thm:pure-merging} can always be made incoherent since
the protocol is one-way with her final state being incoherent. Thus
our result also applies to the scenario of bipartite local incoherent
operations and classical communications (LICC) \cite{Streltsov2015c,Chitambar2015a}.

\bigskip{}

\begin{figure}
\includegraphics[width=1\columnwidth]{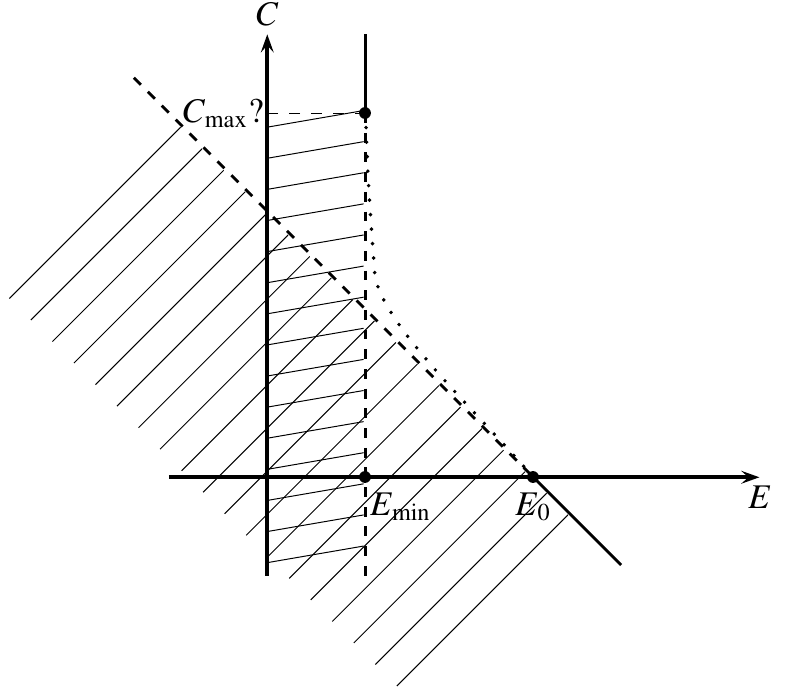} \caption{The achievable region and known bounds for coherence and entanglement
required to merge a general pure state $\psi^{RAB}$. The shaded regions
to the left and below the straight lines are ruled out. The solid
line of slope $-1$ to the right downward from $(E_{0},0)$, as well
as the solid vertical line upward from $(E_{\min},C_{\max})$, are
part of the boundary of the achievable region. The dotted curve connecting
these two points represents $C(E)$, the general form of which is
not known, however. The quantities $E_{0}$ and $E_{\min}$ are given
as $E_{0}=S(A|B)_{\overline{\rho}}$, $E_{\min}=S(A|B)_{\rho}$.}

\label{fig:tradeoff} 
\end{figure}

\textbf{\emph{Coherence-entanglement tradeoff.}} The development so
far revealed some facts about the landscape of the achievable pairs
$(E,C)$ for incoherent merging of a state $\rho^{RAB}$. Most importantly,
there are two inaccessible regions given by the inequalities $E+C\geq S(\Delta^{AB}[\rho])-S(\Delta^{B}[\rho])$
and $E\geq E_{\min}$. For a pure state, these simplify to $E+C\geq S(A|B)_{\overline{\rho}}$
and $E\geq S(A|B)_{\rho}$, and the lower bound is tight as $(E=E_{0}=S(A|B)_{\overline{\rho}},C=0)$
is achievable. Furthermore, since with every achievable pair $(E,C)$,
also $(E+t,C-t)$ is achievable for $t>0$, we find a boundary of
the achievable region in the line of slope $-1$ from $(E_{0},0)$
to the right, see Fig.~\ref{fig:tradeoff}. We do not know at this
point whether this boundary line continues with slope $-1$ also to
the left of that point. The biggest open question is the characterization
of $C_{\max}$, which is the coherence rate required for the minimum
possible entanglement rate $E_{\min}$. Naturally, if we could show
that $(E=E_{\min},C=E_{0}-E_{\min})$ is achievable, we would have
characterized the entire achievable region, showing that it is delimited
by the two above mentioned linear inequalities. On the other hand,
it is quite conceivable that in general $C_{\max}\gg E_{0}-E_{\min}$.

We are now going to present an example indicative of the second option
inspired by the ``flower states'' \footnote{Note that the flower states presented in \cite{Horodecki2005b} are
obtained by purifying the information locking states of \cite{DiVincenzo2004},
and their results are mathematically equivalent due to the duality
between entanglement of formation and Henderson-Vedral information.}: 
\begin{equation}
\begin{split}\ket{\psi}^{RAB} & =\frac{1}{\sqrt{2d}}\sum_{i=0}^{1}\sum_{j=1}^{d}(U_{i}^{\top}\ket{j})^{R}\ket{i}^{A}\ket{j}^{B}\\
 & =\frac{1}{\sqrt{2}}\left(\ket{0}^{A}\otimes(\openone\otimes U_{0})\ket{\Phi_{d}}^{RB}+\ket{1}^{A}\otimes(\openone\otimes U_{1})\ket{\Phi_{d}}^{RB}\right),
\end{split}
\label{eq:flower}
\end{equation}
where for definiteness $U_{0}=\openone$, $U_{1}=\text{QFT}$ is the
quantum Fourier transform, and $\ket{\Phi_{d}}=\sum_{i}\ket{ii}/\sqrt{d}$
is the maximally entangled state. One checks that for this family
of states, $E_{0}=1$ (attained by simply teleporting Alice's qubit)
and $E_{\min}=0$. Indeed, there is a simple exact merging protocol
not using any entanglement, which consists of Alice measuring in the
computational basis and communicating $i$ to Bob. Bob in turn applies
$U_{i}^{\dagger}$ after which he is left with the maximally entangled
state $\ket{\Phi_{d}}^{RB}$ with the reference; now he creates the
state $\ket{+}^{\tilde{B}}=\frac{1}{\sqrt{2}}(\ket{0}+\ket{1})$ and
recovers the state $\ket{\psi}^{R\tilde{B}B}$ by the controlled unitary
$\proj{0}\otimes U_{0}+\proj{1}\otimes U_{1}$. Note that while $U_{0}$
is trivial, $U_{1}$ requires a large amount of coherence to be implemented,
indeed, the previous procedure of Bob requires asymptotically a rate
of $1+\frac{1}{2}\log d$ of coherence. Conversely, we have the following
lower bound: 
\begin{thm}
\label{thm:tradeoff}Merging the state in Eq.~(\ref{eq:flower})
via one-way LQICC without any initial entanglement, i.e.~not only
$E_{i}=0$ but also $\delta=0$ in Eq.~(\ref{eq:achievable}), requires
a rate of coherence at least $C\geq1+\frac{1}{2}\log d\gg1$. 
\end{thm}
\noindent We refer to Appendix \ref{sec:Theorem3} for the proof.
While we proved the theorem for the case where classical communication
only goes in one direction, it is reasonable to believe that this
result can be extended to arbitrary LQICC protocols. We also note
another limitation of the result: Our proof covers only the case that
entanglement is exactly zero initially. It is not clear if this result
also applies when considering more general merging procedure where
entanglement vanishes only in the asymptotic limit. Nevertheless,
this result provides strong evidence that in the task of quantum state
merging it is possible to save a large amount of local coherence by
using little extra entanglement.

\bigskip{}

\textbf{\emph{Application: A family of separable states.}} We will
now apply the results presented so far to the following family of
states: 
\begin{equation}
\rho=\sum_{i,j}p_{ij}\proj{ij}^{R}\otimes\proj{\psi_{ij}}^{A}\otimes\proj{i}^{B},
\end{equation}
where the states $\ket{\psi_{ij}}$ are mutually orthogonal for different
$j$, i.e., $\braket{\psi_{ij}|\psi_{ik}}=\delta_{jk}$. As is shown
in Appendix \ref{sec:proof-separable}, for this type of states all
optimal pairs are given by 
\begin{equation}
(E,C)=(aC_{\max},[1-a]C_{\max})\label{eq:example}
\end{equation}
with $a\geq0$ and $C_{\max}=\sum_{i,j}p_{ij}S(\Delta(\psi_{ij}))$.

\bigskip{}

\textbf{\emph{Conclusions}}\textbf{.} In the present paper we introduced
and studied the task of incoherent quantum state merging. This task
is the same as standard quantum state merging, up to the fact that
one of the parties has free access to local incoherent operations
only, and has to consume a coherent resource for more general operations.
The amount of resources needed for merging is quantified by an entanglement-coherence
pair $(E,C)$. In general, we showed that the entanglement-coherence
sum $E+C$ is nonnegative, which means that no merging procedure can
gain entanglement and coherence at the same time. For pure states
we gave a protocol of incoherent quantum state merging by finding
the minimal entanglement-coherence sum $E+C$, which turns out to
be the conditional entropy of the decohered state $\overline{\rho}^{AB}$.

Our results include an incoherent version of Schumacher compression.
In particular, if we require that the decompression is performed via
incoherent operations only, then the optimal compression rate is given
by $S(\Delta(\rho))$. This rate is in general larger than the standard
compression rate $S(\rho)$, which comes from the fact that coherence
is required for the decompression in the standard case.

We have also made first steps towards an understanding of the precise
tradeoff between entanglement and coherence for the task of LQICC
merging. While this remains a major open problem in general, we have
given strong indications that in certain situations the equivalent
of one ebit can be an arbitrary amount of coherence, which we could
prove in a setting of one-way LQICC and a situation where we want
to reduce the available entanglement exactly (and not only asymptotically)
to zero.

Another open question is the relation of LQICC merging to the results
presented in \cite{Rio2011}. In particular, the authors of \cite{Rio2011}
study the work cost for erasing a system $A$ which is (quantum) correlated
with another observer $B$ in an environment at temperature $T$.
As was shown in \cite{Rio2011}, this work cost is bounded above by
$S(A|B)kT\ln(2)$, where $k$ is the Boltzmann constant. At this point
it is natural to ask if our results can be applied to understand the
role of coherence in the erasure process. We leave these questions
for future research.

\bigskip{}

\textbf{\emph{Acknowledgements.}} We thank Martin B. Plenio for comments
on the manuscript. We acknowledge financial support from the Alexander
von Humboldt Foundation, the John Templeton Foundation, the European
Commission (QUIC and STREP RAQUEL), the European Research Council
(AdG OSYRIS and AdG IRQUAT), the US National Science Foundation (CAREER
1352326), the Spanish MINECO (FIS2013-46768, FIS2008-01236, and FIS2013-40627-P)
with the support of FEDER funds, the Generalitat de Catalunya (2014-SGR-874
and 2014-SGR-966), and Fundació Privada Cellex.

 \bibliographystyle{apsrev4-1}
\bibliography{literature}

\appendix

\section{\label{sub:Proposition1}Proof of Proposition \ref{prop:bound-1}}

Here we will prove that any achievable pair $(E,C)$ fulfills the
following inequality: 
\begin{equation}
E+C\geq S\left(\Delta^{AB}\left[\rho\right]\right)-S\left(\Delta^{B}\left[\rho\right]\right).
\end{equation}

For proving this statement, we will use the QI relative entropy which
can be written as \cite{Chitambar2015}: 
\begin{equation}
C_{r}^{X|Y}\left(\rho^{XY}\right)=S\left(\Delta^{Y}\left[\rho^{XY}\right]\right)-S\left(\rho^{XY}\right).
\end{equation}
The definition of an achievable pair in Eq.~(\ref{eq:achievable})
of the main text together with the continuity of the QI relative entropy
\cite{Chitambar2015} implies that there exist nonnegative numbers
$E_{i}$, $E_{t}$, $C_{i}$, and $C_{t}$ with $E=E_{i}-E_{t}$ and
$C=C_{i}-C_{t}$ such that for any $0<\varepsilon\leq1/2$ and any
$\delta>0$ there is an integer $n\geq1$ and an LQICC protocol $\Lambda$
such that 
\begin{align}
 & C_{r}^{RA|B\tilde{B}}\left(\Lambda\left[\rho_{i}^{\otimes n}\otimes\Phi_{2}^{\otimes\left\lfloor (E_{i}+\delta)n\right\rfloor }\otimes\Psi_{2}^{\otimes\left\lfloor (C_{i}+\delta)n\right\rfloor }\right]\right)\geq\label{eq:proofProp1}\\
 & C_{r}^{RA|B\tilde{B}}\left(\rho_{t}^{\otimes n}\otimes\Phi_{2}^{\otimes\left\lceil E_{t}n\right\rceil }\otimes\Psi_{2}^{\otimes\left\lceil C_{t}n\right\rceil }\right)-2\varepsilon\log_{2}d_{\mathrm{tot}}-2h(\varepsilon).\nonumber 
\end{align}
Here, $h(x)=-x\log_{2}x-(1-x)\log_{2}(1-x)$ is the binary entropy
and $d_{\mathrm{tot}}$ is the total dimension given as follows: 
\begin{equation}
d_{\mathrm{tot}}=d_{RA\tilde{B}B}^{n}\times4^{\left\lfloor (E_{i}+\delta)n\right\rfloor +\left\lceil E_{t}n\right\rceil }\times2^{\left\lfloor (C_{i}+\delta)n\right\rfloor +\left\lceil C_{t}n\right\rceil }.
\end{equation}
In the next step we will introduce the number $d'$ as follows: 
\begin{equation}
d'=d_{RA\tilde{B}B}\times4^{E_{i}+E_{t}+\delta+1}\times2^{C_{i}+C_{t}+\delta+1},
\end{equation}
and it can be verified by inspection that $(d')^{n}\geq d_{\mathrm{tot}}$.
Together with Eq.~(\ref{eq:proofProp1}) this leads us to the following
inequality: 
\begin{align}
 & C_{r}^{RA|B\tilde{B}}\left(\Lambda\left[\rho_{i}^{\otimes n}\otimes\Phi_{2}^{\otimes\left\lfloor (E_{i}+\delta)n\right\rfloor }\otimes\Psi_{2}^{\otimes\left\lfloor (C_{i}+\delta)n\right\rfloor }\right]\right)\geq\\
 & C_{r}^{RA|B\tilde{B}}\left(\rho_{t}^{\otimes n}\otimes\Phi_{2}^{\otimes\left\lceil E_{t}n\right\rceil }\otimes\Psi_{2}^{\otimes\left\lceil C_{t}n\right\rceil }\right)-2n\varepsilon\log_{2}d'-2h(\varepsilon).\nonumber 
\end{align}
Since the QI relative entropy is additive and does not increase under
LQICC operations \cite{Chitambar2015}, it follows that 
\begin{align}
 & C_{r}^{RA|B\tilde{B}}\left(\rho_{i}\right)+\frac{\left\lfloor (E_{i}+\delta)n\right\rfloor +\left\lfloor (C_{i}+\delta)n\right\rfloor }{n}\\
 & \geq C_{r}^{RA|B\tilde{B}}\left(\rho_{t}\right)+\frac{\left\lceil E_{t}n\right\rceil +\left\lceil C_{t}n\right\rceil }{n}-2\varepsilon\log_{2}d'-\frac{2}{n}h(\varepsilon).\nonumber 
\end{align}
The desired statement follows by using the relations: 
\begin{align}
C_{r}^{RA|B\tilde{B}}\left(\rho_{i}\right) & =S(\Delta^{B}(\rho))-S(\rho),\\
C_{r}^{RA|B\tilde{B}}(\rho_{t}) & =S(\Delta^{AB}(\rho))-S(\rho)
\end{align}
together with the facts that $\left\lfloor x\right\rfloor \leq x$
and $\left\lceil x\right\rceil \geq x$.

Note that in this proof, the restriction to LQICC operations is needed
only to ensure monotonicity of the QI relative entropy. This monotonicity
follows from the fact that LQICC operations preserve the set of so-called
quantum-incoherent QI states; i.e. states of the form $\rho^{XY}=\sum_{y}p_{y}\rho_{y}^{X}\otimes\proj{y}^{Y}$,
where the $\rho_{y}^{X}$ are arbitrary and $\ket{y}$ is the incoherent
basis for system $Y$. The QI relative entropy is therefore a monotone
for any other class of operations that also preserve the QI set of
states.

The most general class of QI-preserving operations are formed by so-called
``maximal'' incoherent operations (MIO) on Bob's side and arbitrary
operations on Alice's. Recall that a completely positive trace-preserving
(CPTP) map $\mathcal{E}$ belongs to the class MIO if $\mathcal{E}(\rho)\in\mathcal{I}$
for any $\rho\in\mathcal{I}$, where $\mathcal{I}$ denotes the set
of incoherent states. A MIO measurement that produces classical outcomes
$i$ can be represented by the CPTP map $\rho^{B}\mapsto\sum_{i}\mathcal{E}_{i}(\rho)^{B}\otimes\proj{i}^{C}$
with each $\mathcal{E}_{i}$ being an incoherent CP map. Hence the
incoherent operations $\{K_{i}\}_{i}$ studied in this paper are special
MIO maps of the form $\rho\mapsto\sum_{i}K_{i}\rho K_{i}^{\dagger}\otimes\proj{i}$.
It is easy to see that MIO acts invariantly on the set of QI states.
Thus, the QI relative entropy is monotonic under MIO, and we see that
Proposition \ref{prop:bound-1} also holds for MIO performed on Bob's
side.

\section{\label{sec:Theorem2}Proof of Theorem \ref{thm:pure-merging}}

Here we will prove that any pure state $\ket{\psi}^{RAB}$ can be
merged with the pair $(E_{0},C=0)$, where 
\begin{equation}
E_{0}=S(A|B)_{\overline{\rho}}=S\left(\overline{\rho}^{AB}\right)-S\left(\overline{\rho}^{B}\right),
\end{equation}
and $\overline{\rho}^{X}=\Delta^{X}[\rho^{X}]$ denotes the dephased
state.

For proving this, note that any pure state $\ket{\psi}^{RAB}$ can
be written in the following form: 
\begin{equation}
\ket{\psi}^{RAB}=\sum_{x,y}a_{xy}\ket{\mu_{xy}}^{R}\otimes\ket{x}^{A}\otimes\ket{y}^{B}
\end{equation}
with complex coefficients $a_{xy}$ and arbitrary referee's states
$\ket{\mu_{xy}}^{R}$. We will now show that the state merging transformation
$\ket{\psi}^{RAB}\to\ket{\psi}^{RB'B}$ can performed asymptotically
by LQICC at an entanglement consumption rate of $S(A|B)_{\overline{\rho}}=S\left(\overline{\rho}^{AB}\right)-S\left(\overline{\rho}^{B}\right)$.

By the structure of the state $\ket{\psi}^{RAB}$, we see that $S(A|B)_{\overline{\rho}}=H(X|Y)$,
where $X$ and $Y$ are random variables jointly distributed according
to $p(x,y)=|a_{xy}|^{2}$. The state merging protocol is essentially
Slepian-Wolf data compression \cite{Slepian1973} of the source $X$
with side information $Y$ at the decoder, run in coherent superposition.
The resulting protocol will turn out to be fully incoherent, for both
Alice and Bob.

To be precise, fix a code for block length $n$, consisting of compression
and decompression functions 
\begin{align}
f:\mathcal{X}^{n} & \longrightarrow[N],\\
g:[N]\times\mathcal{Y}^{n} & \longrightarrow\mathcal{X}^{n},
\end{align}
such that $\log N=n\bigl(H(X|Y)+\delta\bigr)$ and 
\begin{equation}
\Pr\left\{ X^{n}\neq g\bigl(f(X^{n}),Y^{n}\bigr)\right\} \leq\epsilon.
\end{equation}
By the Slepian-Wolf theorem~\cite{Slepian1973,Cover2006}, for every
$\epsilon$, $\delta>0$, such a code exists for all sufficiently
large $n$. For the purposes of the quantum protocol, define $G(\nu,y^{n}):=\bigl(g(\nu,y^{n}),y^{n}\bigr)$,
such that 
\begin{equation}
\Pr\left\{ (X^{n},Y^{n})\neq G\bigl(f(X^{n}),Y^{n}\bigr)\right\} \leq\epsilon.
\end{equation}
Because of this, there exists a subset $\mathcal{S}\subset\mathcal{X}^{n}\times\mathcal{Y}^{n}$
such that $\Pr\left\{ (X^{n},Y^{n})\in\mathcal{S}\right\} \geq1-\epsilon$
and $(x^{n},y^{n})=G(f(x^{n}),y^{n})$ for all $(x^{n},y^{n})\in\mathcal{S}$.
We can therefore introduce the one-to-one function $\widetilde{G}:[N]\times\mathcal{Y}^{n}\longrightarrow\mathcal{X}^{n}\times\mathcal{Y}^{n}\stackrel{.}{\cup}\mathcal{R}$
(with $\mathcal{R}=[N]\times\mathcal{Y}^{n}$) by 
\begin{equation}
\widetilde{G}(\nu,y^{n})=\begin{cases}
G(\nu,y^{n}) & \text{ if }G(\nu,y^{n})\in\mathcal{S},\\
(\nu,y^{n}) & \text{ otherwise}.
\end{cases}
\end{equation}
Note that by construction 
\[
\Pr\left\{ (X^{n},Y^{n})\neq\widetilde{G}\bigl(f(X^{n}),Y^{n}\bigr)\right\} \leq\epsilon.
\]

Now we can describe the quantum protocol: Define the incoherent(!)
isometries 
\begin{equation}
U:\ket{x^{n}}^{A^{n}}\longmapsto\ket{x^{n}}^{A^{n}}\ket{f(x^{n})}^{A_{0}},
\end{equation}
and 
\begin{equation}
V:\ket{\nu}^{B_{0}}\ket{y^{n}}^{B^{n}}\longmapsto\ket{\widetilde{G}(\nu,y^{n})}^{(B_{0}+{A'}^{n})B^{n}}.
\end{equation}
The first three steps of the protocol are easy: Alice applies $U$
to her register $A^{n}$, then sends the register $A_{0}$ to Bob
by teleportation \cite{Bennett1993} using $\log N=n(E+\delta)$ ebits,
who receives it in his register $B_{0}$ and applies $V$ to $B_{0}B^{n}$.
The resulting state is 
\begin{equation}
\begin{split} & \ket{\varphi}^{R^{n}A^{n}(B_{0}+{A'}^{n})B^{n}}=(I^{R^{n}}\otimes VI^{A_{0}\rightarrow B_{0}}U)\ket{\psi}^{\otimes n}\\
 & =\sum_{x^{n},y^{n}}a_{x^{n}y^{n}}\ket{\mu_{x^{n}y^{n}}}^{R^{n}}\otimes\ket{x^{n}}^{A^{n}}\otimes\ket{\widetilde{G}(\nu,y^{n})}^{(B_{0}+{A'}^{n})B^{n}}\\
 & =\sum_{(x^{n},y^{n})\in\mathcal{S}}a_{x^{n}y^{n}}\ket{\mu_{x^{n}y^{n}}}^{R^{n}}\otimes\ket{x^{n}}^{A^{n}}\otimes\ket{x^{n},y^{n}}^{(B_{0}+{A'}^{n})B^{n}}\\
 & +\sum_{(x^{n},y^{n})\not\in\mathcal{S}}a_{x^{n}y^{n}}\ket{\mu_{x^{n}y^{n}}}^{R^{n}}\otimes\ket{x^{n}}^{A^{n}}\otimes\ket{v,y^{n}}^{(B_{0}+{A'}^{n})B^{n}}.
\end{split}
\end{equation}
Note that the overall amplitude of the second summation is $<\epsilon$
since $\Pr(\mathcal{S})\geq1-\epsilon$. Furthermore, the $\ket{v}$
are orthogonal to the $\ket{x^{n}}$. Thus by defining 
\begin{equation}
\ket{\widetilde{\psi}}^{RAA'B}:=\sum_{x,y}a_{xy}\ket{\mu_{xy}}^{R}\otimes\ket{x}^{A}\otimes\ket{x}^{A'}\ket{y}^{B},
\end{equation}
we have 
\begin{equation}
\tr\widetilde{\psi}^{\otimes n}\varphi\geq1-\epsilon,
\end{equation}
by the Slepian-Wolf property of the maps $f$ and $\widetilde{G}$.
In other words, 
\begin{equation}
\ket{\varphi}=\sqrt{1-\epsilon}\ket{\widetilde{\psi}}^{\otimes n}+\sqrt{\epsilon}\ket{\theta},\label{eq:almost-there}
\end{equation}
with a (sub-)normalized vector $\ket{\theta}$. It remains to decouple
the register $A^{n}$, which however is easily done due to the structure
of $\ket{\widetilde{\psi}}$ as a generalized GHZ-state: Indeed, for
$d=|\mathcal{X}|=|A|$, consider the conjugate basis 
\[
\ket{\alpha}=\frac{1}{\sqrt{d}}\sum_{x=0}^{d-1}e^{2\pi i\alpha x/d}\ket{x},
\]
then one can confirm by direct calculation 
\begin{align}
\bra{\alpha}\widetilde{\psi}\rangle & =\frac{1}{\sqrt{d}}\sum_{x,y}a_{xy}e^{-2\pi i\alpha x/d}\ket{xy}^{R}\otimes\ket{x}^{A'}\ket{y}^{B}\nonumber \\
 & =\frac{1}{\sqrt{d}}(I_{R}\otimes Z^{-\alpha}\otimes I_{B})\ket{\psi}^{RA'B},
\end{align}
and thus 
\begin{equation}
\bra{\alpha^{n}}\widetilde{\psi}\rangle^{\otimes n}=\frac{1}{d^{n/2}}\bigotimes_{i=1}^{n}(I_{R_{i}}\otimes Z^{-\alpha_{i}}\otimes I_{B_{i}})\ket{\psi}^{R_{i}A_{i}'B_{i}}.
\end{equation}
I.e., to transform $\bigl(\widetilde{\psi}^{RAA'B}\bigr)^{\otimes n}$
to $\bigl(\psi^{RA'B}\bigr)^{\otimes n}$, Alice will destructively
measure each of the $n$ $A$-systems in the conjugate basis $\{\ket{\alpha}\}$
-- which is an incoherent operations -- then communicates the outcomes
$\alpha^{n}=\alpha_{1}\ldots\alpha_{n}$ to Bob who applies the diagonal
unitaries $Z^{\alpha_{i}}$ (to the $i$-th $A'$-system). To be precise,
the incoherent operation that Alice performs is given by Kraus operators
$K_{\alpha^{n}}=\ket{0}\bra{\alpha^{n}}$, which map her system to
the incoherent state $\ket{0}$ for every outcome $\alpha^{n}$. By
applying this procedure to $\varphi$ instead, they obtain a final
state $\psi^{(n)}$ with 
\begin{equation}
\bigl\|\psi^{(n)}-(\psi^{RA'B})^{\otimes n}\bigr\|_{1}\leq\bigl|\ket{\psi^{(n)}}-\ket{\psi}^{\otimes n}\bigr|_{2}\leq2\sqrt{\epsilon}.
\end{equation}
As $\epsilon$ can be made arbitrarily small for increasing $n$,
this concludes the proof.

Note that in this protocol Bob simply performs permutations and diagonal
unitaries. Since these operations belong to the class of physical
incoherent operations (PIO) \cite{Chitambar2016} and the more general
class of strictly incoherent operations (SIO) \cite{Yadin2015b},
we see that Theorem \ref{thm:pure-merging} also holds when Bob is
restricted to PIO/SIO.

\section{\label{sec:Theorem3}Proof of Theorem \ref{thm:tradeoff}}

We now show that for states of the form 
\begin{equation}
\begin{split}\ket{\psi}^{RAB} & =\frac{1}{\sqrt{2d}}\sum_{i=0}^{1}\sum_{j=1}^{d}(U_{i}^{\top}\ket{j})^{R}\ket{i}^{A}\ket{j}^{B}\\
 & =\frac{1}{\sqrt{2}}\left(\ket{0}^{A}\otimes(\openone\otimes U_{0})\ket{\Phi_{d}}+\ket{1}^{A}\otimes(\openone\otimes U_{1})\ket{\Phi_{d}}\right),
\end{split}
\label{eq:flower-1}
\end{equation}
any \emph{one-way} LQICC protocol that does not use any entanglement,
requires a rate of coherence of at least $C'\geq1+\frac{1}{2}\log d\gg1$.
Namely, to succeed in merging of $\psi^{\otimes n}$, Alice's measurement
needs to leave Bob approximately with a maximally entangled state
with $R$, a state $\ket{\phi}^{R^{n}B^{n}}$ within trace distance
$\epsilon$ from $(\openone\otimes V)\ket{\Phi_{d}}^{\otimes n}$,
with a unitary $V$ on $B^{n}$ (at least for most outcomes). Bob's
local operation $T$ then must take $\phi$ to within $\epsilon$
of $\left(\ket{\psi}^{R\tilde{B}B}\right)^{\otimes n}$, which means
that $T$ in a certain precise sense has to approximate the action
of $U^{\otimes n}V^{\dagger}$, with the isometry $U=\frac{1}{\sqrt{2}}\bigl(\ket{0}^{\tilde{B}}\otimes U_{0}+\ket{1}^{\tilde{B}}\otimes U_{1}\bigr)$:
\begin{equation}
\begin{split}(\id\otimes T)\phi & \approx\psi^{\otimes n}\\
 & =(\openone\otimes U^{\otimes n}V^{\dagger})(\openone\otimes V)\Phi_{d}^{\otimes n}(\openone\otimes V)^{\dagger}(\openone\otimes U^{\otimes n}V^{\dagger})^{\dagger}\\
 & \approx(\openone\otimes V)\Phi_{d}^{\otimes n}(\openone\otimes V)^{\dagger},
\end{split}
\end{equation}
where the $\approx$ sign means that the respective states are at
trace distance $\leq\epsilon$. Hence we get 
\begin{equation}
\left\Vert (\id\otimes T)\bigl[(\openone\otimes V)\Phi_{d}^{\otimes n}(\openone\otimes V)^{\dagger}\bigr]-\psi^{\otimes n}\right\Vert _{1}\leq2\epsilon,
\end{equation}
which implies 
\begin{equation}
\frac{1}{d^{n}}\sum_{j^{n}=j_{1}\ldots j_{n}}\bigl\| T(\proj{j^{n}})-U^{\otimes n}V^{\dagger}\proj{j^{n}}V(U^{\otimes n})^{\dagger}\bigr\|_{1}\leq2\epsilon,\label{eq:approximate-decoding-action}
\end{equation}
and applying $\Delta$ to each of the $n$ $\tilde{B}B$ systems,
we get 
\begin{equation}
\frac{1}{d^{n}}\sum_{j^{n}=j_{1}\ldots j_{n}}\Bigl\|\Delta^{\otimes n}\bigl(T(\proj{j^{n}})\bigr)-\Delta^{\otimes n}\bigl(U^{\otimes n}V^{\dagger}\proj{j^{n}}V(U^{\otimes n})^{\dagger}\bigr)\Bigr\|_{1}\leq2\epsilon.
\end{equation}
Now, we claim that for every state $\sigma$ on $B^{n}$, 
\begin{equation}
S\left(\Delta^{\otimes n}\bigl(U^{\otimes n}\sigma(U^{\otimes n})^{\dagger}\bigr)\right)\geq n\left(1+\frac{1}{2}\log d\right).\label{eq:entropy-bound}
\end{equation}
This is easy to see for $n=1$, since 
\begin{equation}
M(\sigma):=\Delta(U\sigma U^{\dagger})=\frac{1}{2}\proj{0}\otimes\Delta(U_{0}\sigma U_{0}^{\dagger})+\frac{1}{2}\proj{1}\otimes\Delta(U_{1}\sigma U_{1}^{\dagger}),
\end{equation}
whose entropy is lower bounded for every state by $1+\frac{1}{2}\log d$,
according to the Maassen-Uffink entropic uncertainty relation \cite{Maassen1988};
namely, note that $\Delta(U_{i}\sigma U_{i}^{\dagger})$ correspond
to measuring $\sigma$ in one of two mutually unbiased bases. To obtain
eq.~(\ref{eq:entropy-bound}), we observe that the state of which
we need the entropy is $M^{\otimes n}(\sigma)$, and by the additivity
of the minimum output entropy of entanglement-breaking channels \cite{Shor2002,King2002},
this is at least $n$ times the single-copy bound. Thus, by eq.~(\ref{eq:approximate-decoding-action})
and the asymptotic continuity of $C_{r}$, the relative entropy of
coherence \cite{Winter2015}, we get 
\begin{equation}
\frac{1}{d^{n}}\sum_{j^{n}=j_{1}\ldots j_{n}}C_{r}\bigl(T(\proj{j^{n}})\bigr)\geq n\left(1+\frac{1}{2}\log d\right)-2n\epsilon-2.
\end{equation}
What this says is that $T$ is capable of generating a large amount
of coherence, namely at least $n\left(1+\frac{1}{2}\log d\right)-O(n\epsilon)$.
Clearly, this implies that to implement $T$, Bob must consume at
least that amount of pure coherence, and so in the limit of $n\rightarrow\infty$
and $\epsilon\rightarrow0$, we obtain $C'\geq1+\frac{1}{2}\log d$
for the coherence rate required.

We note that the proof also works for the case where general MIO operations
are performed on Bob's side. Thus, the statement of the theorem also
holds in this scenario.

\section{\label{sec:proof-separable}Proof of Eq.~(\ref{eq:example})}

Here we will consider the following family of states: 
\begin{equation}
\rho=\sum_{i,j}p_{ij}\proj{ij}^{R}\otimes\proj{\psi_{ij}}^{A}\otimes\proj{i}^{B},
\end{equation}
where the states $\ket{\psi_{ij}}$ are mutually orthogonal for different
$j$, i.e., $\braket{\psi_{ij}|\psi_{ik}}=\delta_{jk}$. We will now
show that for these states all optimal pairs are given by 
\begin{equation}
(E,C)=(aC_{\max},[1-a]C_{\max})\label{eq:example-2}
\end{equation}
with $a\geq0$ and $C_{\max}=\sum_{i,j}p_{ij}S(\Delta(\psi_{ij}))$.

For proving this we first invoke Proposition~\ref{prop:bound-1}
in the main text, which implies that any achievable pair is bounded
below by 
\begin{align}
E+C & \geq\sum_{i,j}p_{ij}S(\Delta(\psi_{ij})).\label{eq:example-1}
\end{align}
In the next step note that for this family of states merging is achievable
without entanglement, and thus $E_{\min}=0$. From Eq.~(\ref{eq:example-1})
it follows that 
\begin{align}
C_{\max} & \geq\sum_{i,j}p_{ij}S(\Delta(\psi_{ij})).
\end{align}
Now note that $(0,\sum_{i,j}p_{ij}S(\Delta(\psi_{ij})))$ is an achievable
pair, which can be achieved if Bob performs a von Neumann measurement
in the basis $\ket{i}^{B}$ and communicates the result of the outcome
to Alice. Depending on the outcome $i$ of Bob's measurement, Alice
performs a von Neumann measurement in the basis $\{\ket{\psi_{ij}}\}_{j}$,
and communicates her outcome to Bob. Depending on the outcomes $i$
and $j$, Bob prepares his additional system $\tilde{B}$ in the state
$\ket{\psi_{ij}}^{\tilde{B}}$, and the merging procedure is complete.
Since the coherence cost of preparing the state $\ket{\psi_{ij}}$
is $S(\Delta(\psi_{ij}))$ \cite{Winter2015}, this reasoning proves
that 
\begin{align}
C_{\max} & =\sum_{i,j}p_{ij}S(\Delta(\psi_{ij})).
\end{align}
Due to the facts that $(0,C_{\max})$ is an achievable pair and that
via LQICC operations Alice and Bob can convert a singlet into a maximally
coherent single-qubit state on Bob's side \cite{Chitambar2015}, it
follows that $(aC_{\max},[1-a]C_{\max})$ is also achievable for all
$a\geq0$. Moreover, all these pairs must be optimal due to Eq.~(\ref{eq:example-1}).
It remains to show that all optimal pairs have this form. For this,
note that any optimal pair $(E,C)$ must have coherence $C\leq C_{\max}$,
and thus we can always write $C=[1-a]C_{\max}$. Then, in order for
the pair to be optimal, its entanglement must be $E=aC_{\max}$. In
particular, the pair is not achievable if entanglement is below $aC_{\max}$,
and the pair is not optimal if entanglement is above this value. 
\end{document}